\documentclass[reprint]{revtex4-1}
\usepackage{amsmath,amssymb,graphicx}
\usepackage{graphicx}
\usepackage{dcolumn}
\usepackage{bm}
\usepackage{booktabs}

\begin{document}

\preprint{APS/123-QED}

\title{A new scheme of compact cold atom clock based on diffuse laser cooling in a cylindrical cavity}
\author{Peng Liu, Yanling Meng, Jinyin Wan, Xiumei Wang, Yaning Wang,
Ling Xiao}
\author{Huadong Cheng}
\email{corresponding author: chenghd@siom.ac.cn}
\author{Liang Liu}
\email{corresponding author: liang.liu@siom.ac.cn}
 \affiliation{Key Laboratory of Quantum Optics and Center of Cold Atom Physics, Shanghai Institute of Optics and Fine Mechanics, Chinese Academy of Sciences, Shanghai 201800, China.}
\date{\today}

\begin{abstract}
We present a new scheme of compact Rubidium cold-atom clock which performs the diffuse light cooling, the microwave interrogation and the detection of the clock signal in a cylindrical microwave cavity. The diffuse light is produced by the reflection of the laser light at the inner surface of the microwave cavity. The pattern of injected laser beams is specially designed to make most of the cold atoms accumulate in the center of the microwave cavity. The microwave interrogation of cold atoms in the cavity leads to Ramsey fringes whose line-width is $24.5$~Hz and the contrast of $95.6\%$ when the free evolution time is 20~ms. The frequency stability of $7.3\times10^{-13}\tau^{-1/2}$ has been achieved recently. The scheme of this physical package can largely reduce the complexity of the cold atom clock, and increase the performance of the clock.

\begin{description}
\item[PACS numbers]
37.10.De, 42.50.Ct, 42.62.Fi, 06.20.-f
\end{description}
\end{abstract}

\maketitle


\section{INTRODUCTION}

Over the past 20 years, impressive progress of atomic clocks has been achieved due to the applications of laser and cold atom techniques. Laser pumped Rubidium vapor cell clock has been demonstrated with the replacement of the lamp by a compact frequency-stabilized laser~\cite{Affolderbach06}. Godone et al. developed a clock scheme called pulsed optically-pumped (POP) in a vapor-cell which uses a diode laser for the optical pumping and the Ramsey technique to interrogate atoms~\cite{Godone06IE,Godone06pra,Micalizio12}.  In this scheme, the light shift can be greatly reduced.

With the development of cold atom technique, for POP, the vapor-cell can be replaced by cold atoms with less temperature sensitivity and much longer coherent time which could improve the clock’s performance. M\"{u}ller et al. have cooled and interrogated cesium atoms inside a cylindrical stainless steel cavity where a Magneto-Optical Trap (MOT) was applied~\cite{muller11}. However, its strong magnetic field will lead to the magnetization of stainless steel for long term operation. Consequently, the performance of the clock will be reduced. Besides, 6 windows on the cavity required by MOT also affect the inner microwave field mode.

A more attractive method to cool atoms for a POP clock is diffuse light cooling~\cite{Guillot01,Cheng09,Zhang09}. Typically, diffuse light is generated inside a sphere by diffuse multi-reflection of injected lasers on the inner surface of the sphere. Diffuse light cooling does not require fine adjustment of lasers and can cool atoms efficiently around the zero or weak magnetic fields. Thus this scheme is very robust. The cesium atom clock named HORACE using the three dimensional (3D) isotropic light cooling configuration has achieved a relative frequency stability of $2.2\times10^{-13}\tau^{-1/2}$  and $3\times10^{-15}$  at the integrating time of $10^{4}$~s~\cite{Esnault10,Esnault11}. A better performance is expected in microgravity environment which makes the HORACE a strong candidate for onboard satellite clock. We have developed a similar clock with rubidium atoms~\cite{Zheng13}.

In this paper, we describe the development of a compact cold atom clock based on diffuse light cooling in a cylindrical microwave cavity. Different from the HORACE which uses a spherical cavity, a cylindrical cavity is used in our experiment both for diffuse light cooling and microwave interrogation with cold atoms~\cite{Meng14,Meng14Acta,Meng13}.  Furthermore, a technique to control the distribution of cold atom cloud inside the cavity is applied, by this way, more atoms can be accumulated at the center of the cavity~\cite{Meng13,Meng14pla}.

This paper is organized as follows. Section~\ref{2} gives a detailed description of the optical and physical package. Section~\ref{3} describes the time sequence of the experiment and discusses the frequency stability of the atomic clock. Section~\ref{4} evaluates the factors that affect the frequency stability of the atomic clock. Section~\ref{5} concludes the results and discusses its further applications.

\section{THE CONFIGRATION OF THE RUBIDIUM ATOM CLOCK}\label{2}
\begin{figure}[htbp]
\centerline{\includegraphics[width=0.4\textwidth]{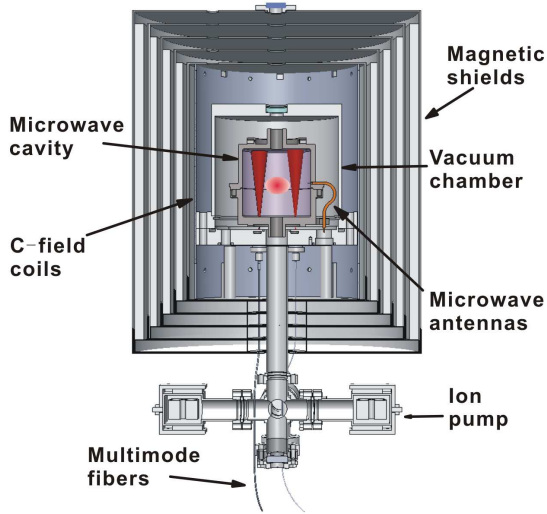}}
\caption{The physical package of the atom clock. The cold atoms are captured in the central region of the microwave cavity and the density of the cold atoms is about 4 times larger than the previous scheme of the old cooling structure~\cite{Zheng13}.}\label{Fig-phy}
\end{figure}

The physical package is shown in Fig.~\ref{Fig-phy}. The microwave cavity is made of oxygen-free copper and the inner surface is coated with silver to increase the diffuse reflectance at 780~nm (96.6\%). Four holes with diameter of 3 mm are centrosymmetric distributed in the undersurface of the cavity for injecting lasers vertically to the microwave cavity. Four multimode fibers are used to transport the cooling, repumping and pumping lights to the physical package~\cite{Meng13}. The imbalance of the light intensities of these four fibers is kept smaller than 0.25\%. For the stage of microwave interrogation, we use a microwave coupling loop horizontally set in the middle of the internal cylinder to pursue large uniform area of microwave field. The microwave cavity operates at the TE011 mode with the quality factor about 11000. We fix the microwave cavity in a vacuum chamber where the vacuum maintains at about $3\times10^{-8}$~Pa by ion pumps. The magnetic coils producing the constant longitudinal magnetic fields (12~mG) are intertwined around the vacuum chamber.  The vacuum chamber and the magnetic coils are covered by five-layer of magnetic shields which can decrease the inside residual stray magnetic field to be smaller than 1~nT. Besides, the first layer is surrounded by the heating coils for the temperature control loop.

\begin{figure}[htbp]
\centerline{\includegraphics[width=0.47\textwidth]{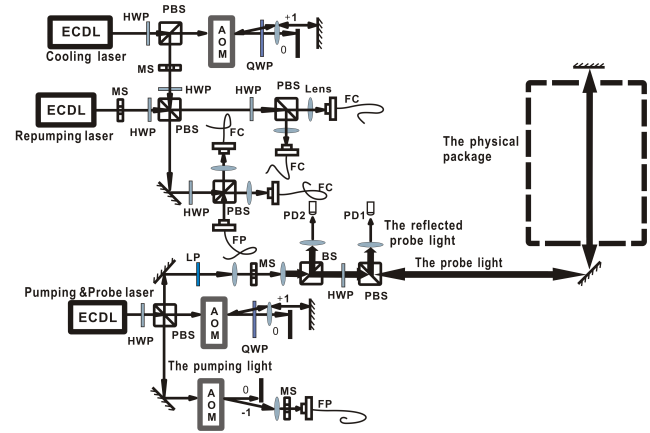}}
\caption{Schematic of the light paths. ECDL-extended-cavity diode laser, QWP-quarter wave plate, HWP-half wave plate, PBS-polarizing beam splitter, BS-beam splitter, LP-linear polarizer, FP- fibers for pumping light, FC-fibers for cooling lights, PD1-photo detector for the detection of clock signals, PD2-photo detector for the intensity stabilizing of the probe light, MS-mechanical shutter.}\label{Fig-light}
\end{figure}

Fig.~\ref{Fig-light} is the schematic of the light paths. Atoms are cooled and manipulated in the center of the cavity~\cite{Meng14pla} with an extended-cavity diode laser (ECDL) as the source of cooling light which is red detuned about 22.8~MHz to the transition $5^{2}S_{1/2}, |F=2\rangle\rightarrow5^{2}P_{3/2}, |F'=3\rangle$ of $^{87}$Rb $D_{2}$ line and a second ECDL which is locked to the transition $5^{2}S_{1/2}, |F=1\rangle\rightarrow5^{2}P_{3/2}, |F'=2\rangle$  provides repumping laser. In the stage of the quantum states preparation, a short pulse of pumping light tuned to the transition of  $5^{2}S_{1/2}, |F=2\rangle\rightarrow5^{2}P_{3/2}, |F'=1\rangle$ is applied. The probe laser emits from the third ECDL with the line-width about 100 KHz whose intensity is stabilized (the intensity fluctuation is detected by the PD2 in Fig.~\ref{Fig-light}) and frequency is shifted near resonant to the transition $5^{2}S_{1/2}, |F=2\rangle\rightarrow5^{2}P_{3/2}, |F'=3\rangle$. The diameter of the probe beam is extended to 6 mm and collimated by two lenses before injecting into the physical package. Since the AOM is not an ideal optical switch and the residual lights will decrease the signal to noise ratio (SNR) of the clock signal and lead to light shift, four low vibration mechanical shutters are used for switching off the cooling, pumping, repumping and probe beams.

\section{THE TIME SEQUENCE OF THE CLOCK OPERATION}\label{3}

\begin{figure}[htbp]
\centerline{\includegraphics[width=0.45\textwidth]{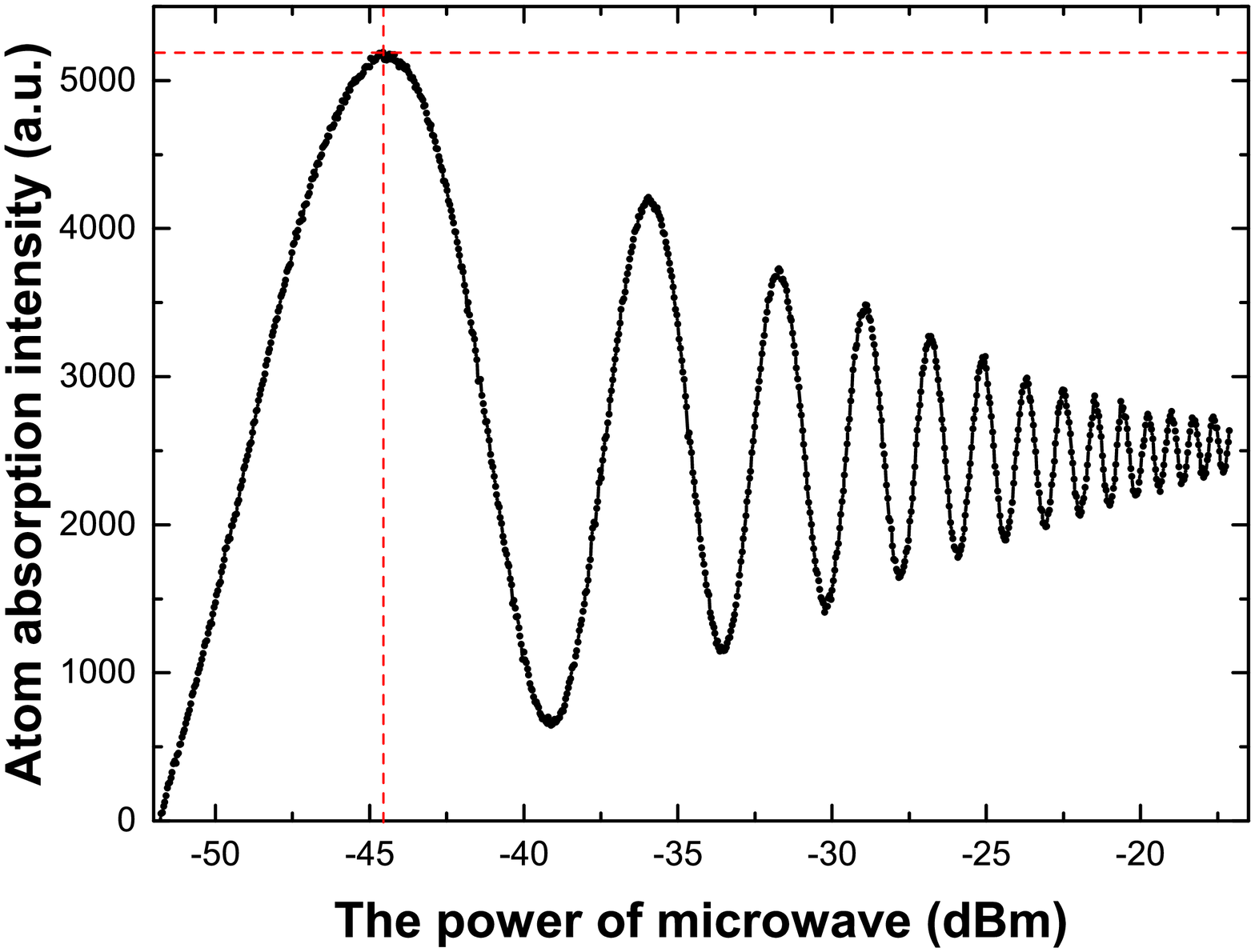}}
\caption{The Rabi oscillation with the microwave pulse duration of 6~ms. Considering the speed and collisions of the cold atoms, the amplitude of the oscillation gets smaller with increasing the power of microwave pulse.}\label{Fig-rabi}
\end{figure}

Typically, the Allan deviation which is used to evaluate the frequency stability of an atomic clock can be written as~\cite{Vanier89}
\begin{equation}\label{sigma}
\sigma_{y}(\tau)=\frac{1}{\pi}\frac{\Delta\nu}{\nu_{0}}\frac{1}{C}\frac{1}{SNR}\sqrt{\frac{T_{c}}{\tau}},
\end{equation}
where $C$ is the contrast of the Ramsey fringes, $\Delta\nu$ is the line-width of the Ramsey fringes and $\tau$ is the integrating time. For the integrating sphere atom clock, the free evolution time $T_{f}$ for the microwave interrogation is mainly limited by the free falling process due to the gravity. Considering the size of the system, the maximum free evolution time on earth is about 50~ms. According to the relationship of $\Delta\nu\sim1/2T_{f}$, longer $T_{f}$  means smaller $\Delta\nu$  while the SNR and contrast of the Ramsey fringes will decrease at the mean time. Thus we need to optimize the parameters of the lasers and the time sequence for operating the atom clock. As for the experiment, to make sure the microwave pulse can realize the largest inversion of the atomic number between the two  $m_{F}=0$ states of the ground states, the Rabi oscillation of the cold atoms needs to be proceeded to find the suitable microwave power. Fig.~\ref{Fig-rabi} shows the Rabi oscillation for the microwave pulse duration of 6~ms by scanning the microwave power. The first peak in Fig.~\ref{Fig-rabi} corresponds to $\Omega t_{m}\approx\pi/2$, where $\Omega$ is the Rabi frequency and $t_{m}=3$~ms.

\begin{figure}[htbp]
\centerline{\includegraphics[width=0.42\textwidth]{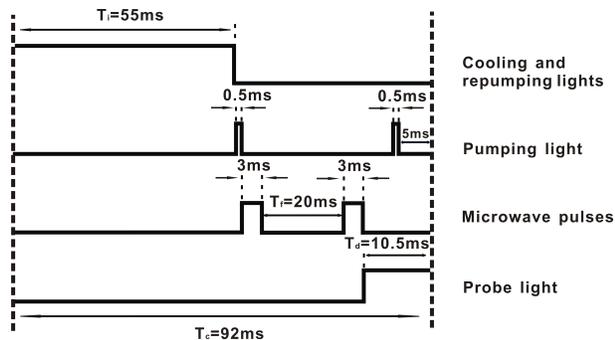}}
\caption{Time sequence of the atomic clock. The cycle time $T_{c}$ is 92~ms. The isotropic laser cooling stage $T_{i}$ lasts 55~ms and the free evolution time $T_{f}$ is 20~ms.}\label{Fig-time}
\end{figure}

Generally, as shown in Fig.~\ref{Fig-time}, the time sequence of a cold atom clock operation includes four stages: cooling of atoms, optical pumping, microwave interrogation and the detection of the clock signal. The 3D isotropic laser cooling lasts 55~ms during which at least $2.4\times10^{8}$ atoms are cooled and accumulated in the central region of the microwave cavity. After the first 0.5~ms optical pumping pulse, atoms are pumped into the $5^{2}S_{1/2}, |F=1\rangle$ state. Then two 3~ms microwave pulses separated by a 20~ms free evolution time are taken for the Ramsey interrogation. After that, the population of the atoms at $5^{2}S_{1/2}|F=2,m_{F}=0\rangle$ state is detected by using a probe beam which is vertical and retroreflected set in the axis of the cylinder cavity. The probe beam lasts 10.5~ms and its intensity is 2.0~$\mu$W. There is a second 0.5~ms optical pumping pulse in the detection window for recording the probe laser intensity. The absorption of the cold atoms is the difference between the two probe transmission signals recorded before and after the second pumping pulse.

\begin{figure}[htbp]
\centerline{\includegraphics[width=0.45\textwidth]{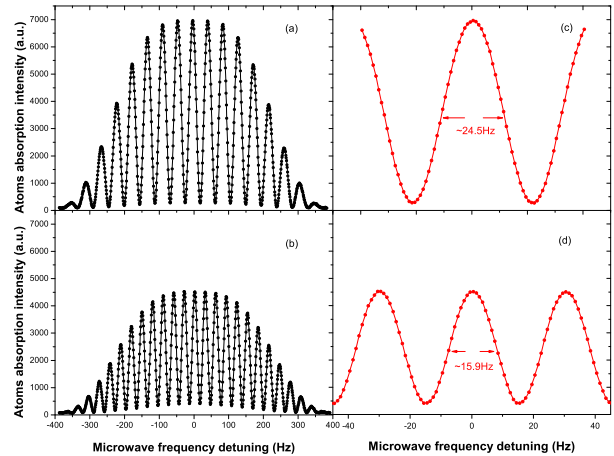}}
\caption{Ramsey fringes with single measurement. (a) The free evolution time $T_{f}$ =20 ms, (b) $T_{f}$=30 ms, (c) shows the central fringes of (a), (d) shows the central fringes of (b).}\label{Fig-ramsey}
\end{figure}

Fig.~\ref{Fig-ramsey} shows the Ramsey fringes with $T_{f}=20$~ms and 30~ms, respectively. It is seen that the line-width of the central fringe changes from 24.5~Hz to 15.9~Hz, and the SNR decreases from about 400 to 260. This is a trade-off between large SNR and narrow line-width.

\section{THE PERFORMANCE OF THE ATOM CLOCK AND THE NOISE ESTIMATION}\label{4}
\begin{figure}[htbp]
\centerline{\includegraphics[width=0.45\textwidth]{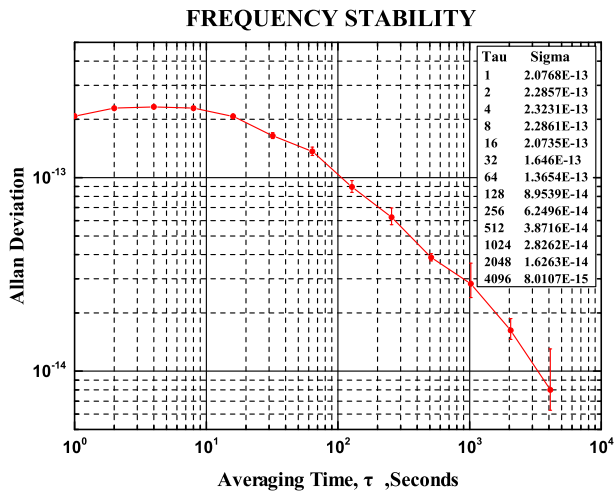}}
\caption{Allan deviation of the cold atomic clock by comparing with the H-maser.}\label{Fig-Allan}
\end{figure}

The error signal to feedback the Local Oscillator (LO) comes from the frequency-hopping detection of the clock transition at the full width at half maximum (FWHM) of the central fringe, and the LO which is the reference source of the microwave frequency synthesizer chain is locked to the central fringe. The relative frequency stability is measured by comparing the frequency output of the LO with an H-maser. Recent result is shown in Fig.~\ref{Fig-Allan}. The frequency stability of $8.01\times10^{-15}$ is reached after integrating of 4000~s.

The main limitations for the short term frequency stability of the clock are the atomic shot noise, the intensity and frequency fluctuations of the probe laser (AM and PM), electronic noise and the local oscillator phase noise (Dick effect). In each cycle, according to the absorption signal, the number of cold atoms detected by the probe light can be estimated with the equation given by~\cite{McDonald09}
\begin{equation}\label{N}
N=\frac{\pi r^{2}}{\sigma_{0}}[\frac{I_{0}-I_{t}}{I_{s}}+(\frac{4\delta^{2}}{\Gamma^{2}}+1)\ln(\frac{I_{0}}{I_{t}})],
\end{equation}
Where $r$ is the radius of the probe beam and $\delta$ is the detuning from atomic resonance. $I_{0}$  and $I_{t}$  are the probe light intensity before and after going through the physical package, respectively. $\Gamma$ is the natural line-width of the transition. $\sigma_{0}$  is the resonant cross-sectional area and $I_{s}$ is the on-resonance saturation intensity. They obey the following relationship~\cite{Steck}
\begin{equation}\label{sigma0}
\sigma_{0}=\frac{h\nu\Gamma}{2I_{s}},
\end{equation}
\begin{equation}\label{Is}
I_{s}=\frac{c\varepsilon_{0}h^{2}\Gamma^{2}}{16\pi^{2}|\hat{\varepsilon\cdot\vec{d}}|^{2}},
\end{equation}
Where $\nu$ is the frequency of the probe light, $\varepsilon_{0}$ is the free space permittivity, and $h$ is the Plank constant. $\hat{\varepsilon}$ is the unit polarization vector of the light field and $\vec{d}$ is the atomic dipole moment. Substituting Eqs.~\eqref{sigma0} and \eqref{Is} into Eq.~\eqref{N}, we can get that the number of the cold atoms detected in each cycle is about $4.7\times10^{7}$ and about 8.5\% of these atoms $N_{a}\approx4.0\times10^{6}$ experience the “0-0” transition. So the corresponding atomic shot noise ($\sigma_{N_{a}}\sim1/\sqrt{N_{a}}$) can be estimated as shown in Table~\ref{noise sources}. 
\begin{table}[b]
\renewcommand{\arraystretch}{1.3}
\setlength\tabcolsep{10pt}
\centering  
\caption{The noise sources and its contribution to the short term frequency stability.}\label{noise sources}
\begin{tabular}{lccc}  
\toprule[2pt]
Noise sources &Uncertainty  \\\midrule[1pt]
Probe laser intensity noise &$2.7\times10^{-13}$ \\         
Atomic shot noise &$4.7\times10^{-13}$  \\        
Electronic noise &$1.8\times10^{-13}$  \\
Local oscillator phase noise &$2.68\times10^{-13}$  \\
Total &$7.3\times10^{-13}$  \\ \bottomrule[2pt]
\end{tabular}
\end{table}

Because the process of the microwave interrogation is not continuously and there is a ``dead time" in each cycle, the phase noises from LO and the frequency synthesizer will degrade the performance of the atomic clock. Here, the limit of Dick effect with the commercial oscillator is about $2.68\times10^{-13}$ on the condition that $S_{y}(f)=10^{-29}f^{2}+1.3\times10^{-27}f+1.3\times10^{-26}$ . Table~\ref{noise sources} gives the noises budget of the cold atom clock.

Cavity pulling effect may be the main contribution to the medium-long-term frequency stability, because it depends on the fluctuations of cold atom number and the cavity detuning from temperature fluctuation. In addition, the microwave power fluctuations are also transferred to the clock transition through cavity pulling~\cite{Micalizio10}. The medium-long-term stability may be improved by decreasing the quality factor of the cavity and increasing the temperature stability of the system.
\section{CONCLUSION}\label{5}
We report a new scheme of compact cold atom clock which integrates all the operation stages in a cylindrical microwave cavity. The use of diffuse laser cooling can simplify the structure and increase the reliability of the system because of its insensitivity to the laser polarization and alignment. The frequency stability of $7.3\times10^{-13}\tau^{-1/2}$ , and $8.01\times10^{-15}$ after integrating 4000~s are achieved. We believe that the frequency stability will be further improved after resolving these technical problems. Furthermore, its all-metal framework and better frequency stability in the microgravity environment make it an excellent option for space applications. In addition, this scheme of physical package can also be operated as a kind of high-efficiency pre-cooling equipment for other physical experiments that need further steps of cooling.

\section{ACKNOWLEDGMENTS}{\centering}
This work was Supported by National High Technology Research and Development Program of China (No. 2012AA120702).

\end{document}